\documentclass[prl,aps,twocolumn,superscriptaddress]{revtex4}
\usepackage{graphicx,amssymb,amsmath,color,psfrag}
\usepackage{amsthm}
\usepackage{amsfonts}
\usepackage{algorithmic}
\usepackage{enumerate}
\usepackage{latexsym}

\begin{document}

\title{Electronic Bloch oscillation in bilayer graphene gradient
superlattices}
\author{Hemeng Cheng}
\affiliation{Department of Physics, Beijing Normal University, Beijing 100875, China}
\author{Changan Li}
\affiliation{Department of Physics, Beijing Normal University, Beijing 100875, China}
\author{Tianxing Ma}
\email{txma@bnu.edu.cn}
\affiliation{Department of Physics, Beijing Normal University, Beijing 100875, China}
\affiliation{Beijing Computational Science Research Center, Beijing 100084, China}
\author{Li-Gang Wang}
\email{sxwlg@yahoo.com}
\affiliation{Department of Physics, Zhejiang University, Hangzhou 310027, China}
\author{Yun Song}
\affiliation{Department of Physics, Beijing Normal University, Beijing 100875, China}
\author{Hai-Qing Lin}
\affiliation{Beijing Computational Science Research Center, Beijing 100084, China}

\begin{abstract}
We investigate the electronic Bloch oscillation in bilayer graphene gradient
superlattices using transfer matrix method. By introducing two kinds of
gradient potentials of square barriers along electrons propagation
direction, we find that Bloch oscillations up to terahertz can occur.
Wannier-Stark ladders,  as the counterpart of Bloch oscillation, are obtained as a series of equidistant transmission peaks, and the
localization of the electronic wave function is also signature of Bloch
oscillation. Furthermore, the period of Bloch oscillation decreases linearly
with increasing gradient of barrier potentials.
%and is affected by the interlayer coupling parameter $t'$.
\end{abstract}

\pacs{73.61.Wp, 73.20.At, 73.21.-b }
\maketitle

%\author{Hemeng Cheng,$^{1}$ ,Changan Li,$^{1}$ , Tianxing Ma,$^{1,2,}$\email{txma@bnu.edu.cn} Li-Gang Wang,$^{3,2,\email{sxwlg@yahoo.com}}$ Yun Song,$^{1}$ and Hai-Qing Lin$^{2}$}

%\affiliation{Beijing Computational Science Research Center,
%Beijing 100084, China}

%\date{\April 15,2014}

%\date{2012.02.11}

%\section{\uppercase\expandafter{\romannumeral1}.  Introduction}

Bloch oscillation (BO) describes the periodic motion of charged particles in
crystals when the particles are subjected to a uniform external electric
field\cite{Bloch1928}. In 1928, Bloch and Zener predicted that
an electron in crystals experienced BO both in momentum and the real space%
\cite{NS2005} when a homogeneous static electrical field is applied, which is
known as electronic Bloch oscillation (EBO).
%An electron in crystals experiences BO both in
%momentum and the real space\cite{NS2005} as predicted by Bloch
%and Zener in 1928 when homogeneous static electrical field is applied,
%which is known as electronic Bloch oscillation(EBO).
In the earlier times, this concept has led to a long time controversy\cite%
{Ricca2003} because of the issues that a constant external electronic field
causes an oscillation current while there were no experimental realizations at
that time. EBO is difficult to be observed in regular crystals
because scattering destroys the coherence of Bloch states before a single BO
can be completed\cite{GM2001,AV2005}. The frequency counterpart of BO is the
Wannier-Stark ladders (WSLs), a series of energy levels separated by a
constant value\cite{AG1960}. Later until 1992, the appearance
of semiconductor superlattices, which have a long periodicity $d$ and a
narrow miniband width, makes the observation of EBO and WSLs possible\cite{JF1992,KL1992,CW1993}.
%and %there were experimental observations of EBO.
Since then, BO has been investigated extensively in
semicoductor superlattices both theoretically and experimentally\cite{JF1992,NS2005}.
%, and some other systems\cite{Ricca2003,GM2001,VL2005,VA2004}.}
%, as well as coupled nanocavity structers\cite{ND2010}.
%, and Bose-Einstein condensates in optical
%lattices\cite{MG2001,OM2001}.

Recently, the experimental realization of graphene superlattices\cite{ALVa2008,JC2008} has drawn much attentions\cite{Park2008,TXMa2012,Liang2011,LGWang2010,MB2009,DD2008,VK2012,AR2013,GJ2011,LK2012,ED2014}.
% and interesting properties in
%graphene-based superlattices have been studied intensively.
In graphene,
the low-energy charge carries behave as massless Dirac fermions,
%and the resulting linear energy dispersion relation
which leads to many interesting electronic properties, and the bilayer
graphene (BLG) provides a semiconductor with a gap tunable by the
electric field effect\cite{Novoselov2004,ZY2005}. For BO, it has
important potential applications, such as designing infrared detectors,
emitters, or lasers which can be tuned in the terahertz frequency range
simply by varying the applied electric field\cite{LL2011,KL2003}. This
raises the questions of the existence and %some
interesting properties of BO
in graphene-based superlattices (GSLs). However, only a few works showing
properties of BO related with graphene until now\cite{DD2008,VK2012,AR2013,GJ2011,LK2012,ED2014}, and its experimental realization is still in vacancy.

In this letter, we investigate EBO in bilayer
graphene gradient superlattices, and propose a general
way to control the BO. BLG is an attractive candidate for transistor
applications since it has a tunable gap which varies in proportion to the
electric field perpendicular to the layers\cite{MK2011}. By studying the
band structures, transmission spectrum and wave functions, it is significant
to see that terahertz Bloch oscillations and WSLs can be
generated in BLG GSLs. %We found that
The band structures are spatially
titling and separated by gaps, and WSLs are seen as a series of equidistant
transmission peaks. Moreover, the wave functions localize spatially, which
also works as the evidence of BO.
\begin{figure}[tbp]
\centering
\includegraphics[scale=0.375]{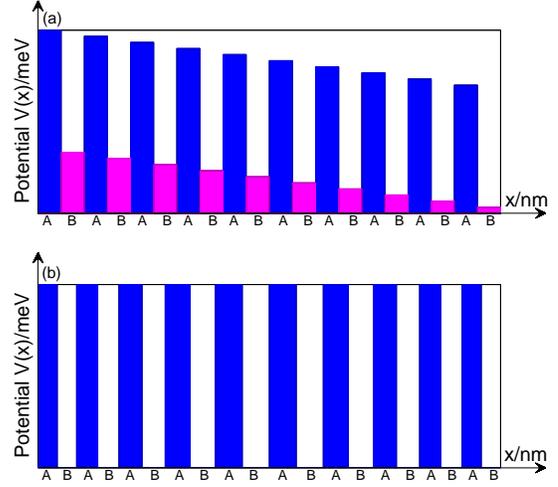}
\caption{(Color online)The potential structures of BLG GSLs. (a)the linear
gradient $\Delta{V}$ in the potential weights of barriers A and B. (b)the
increasing $\Delta{w}$ in the widths of barriers and wells in the front half
and the decreasing $\Delta{w}$ in the rear half of the whole structure.}
\label{Fig:fig1}
\end{figure}

%In particular, our system of BLG GSLs is shown as Fig. \ref{Fig:fig1}.
Fig. \ref{Fig:fig1} shows the %investigated
system of BLG GSLs. % that we studied.
In Fig. \ref{Fig:fig1}(a), we introduce a gradient in the
potential height of higher barriers (A) and lower barriers (B) by decreasing
$\Delta{V}$ per period. Here, A (B) denotes alternating barriers $V_{A}
(V_{B})$ with the widths $w_{A} (w_{B})$. A period contains $V_{A}$ and $%
V_{B}$. At the $i$th period, the potentials with coordinate can be expressed
briefly as $V_{A}(x)$=$V_{A1}-\Delta{V}\times(i-1)$ when $%
\Lambda\times(i-1)\leq x\leq \Lambda\times(i-1)+w_{A}$ and $V_{B}(x)$=$%
V_{B1}-\Delta{V}\times(i-1)$ when $\Lambda\times(i-1)+w_{A}\leq x\leq
\Lambda\times i$, where $1\leq i\leq N$, $N$ is the number of period and $%
\Lambda$=$w_{A}+w_{B}$. In Fig. \ref{Fig:fig1}(b), we introduce a gradient $%
\Delta{w}$ in the widths of barriers (A) and wells (B) while keeping other
parameters constant. In the same period, the width of barrier and well is
the same. At $i$th period, the width of barrier (or well) is $w_{i}$=$%
w_{1}+\Delta{w}\times(i-1)$ when $1\leq i\leq \frac{N}{2}$, and $w_{i}$=$%
w_{N}+\Delta{w}\times(N-i)$ when $\frac{N}{2}\leq i\leq N$. Here the
potential structure is symmetric so $w_{1}$=$w_{N}$. As one will see in Fig. %
\ref{Fig:fig6}, this specific structure can make an interesting and
symmetric band structure.

%We consider the electronic structure of BLG with energy and wave vector
%close to the K point, so the
We consider the %low energy
 one-particle Hamiltonian for BLG %is %given as
\begin{equation}
\hat{H}=\left(
\begin{array}{cccc}
V(x) & \pi & t^{\prime } & 0 \\
\pi^\dag & V(x) & 0 & 0 \\
t^{\prime } & 0 & V(x) & \pi^\dag \\
0 & 0 & \pi & V(x)%
\end{array}
\right),
\end{equation}
where $V(x)$ is the electrostatic potential, $t^{\prime }$ is the interlayer
coupling,  the momentum operators %and defined by
$\pi=-i\hbar\upsilon _{F}\lbrack\frac{\partial }{\partial x}-i%
\frac{\partial}{\partial y}\rbrack$, $\pi^\dag=-i\hbar\upsilon _{F}\lbrack%
\frac{\partial }{\partial x}+i\frac{\partial}{\partial y}\rbrack $, $%
\upsilon _{F}\approx 10^{6}$m/s is the Fermi velocity. This Hamiltonian
satisfies the eigenequation $\hat{H}\Phi$=$E\Phi$, where $\Phi$ is a
four-component spinors $\Phi={\left(\widetilde{\varphi}_{1},\widetilde{%
\varphi}_{2},\widetilde{\varphi}_{3},\widetilde{\varphi}_{4}\right)}^{T}$.
Due to the translation invariance in the $y$ direction, the wave function
can be rewritten as $\widetilde{\varphi}_{m}=\varphi_{m}e^{ik_{y}y}$, $%
m=1,2,3,4$. The solution of eigenequation leads to a transfer matrix%\begin{equation}
%\varphi(x_{j}+\Delta x)=M\varphi(x_{j}),
%\end{equation}
%where the matrix M can be given as\\
\begin{equation}
M_{j}(\Delta x,E,k_{y})=
\begin{pmatrix}
M_{j +} & 0 \\
0 & M_{j -}%
\end{pmatrix}%
\end{equation}
which connects the wave function at any two points inside the $j$th potential, and
\begin{equation}
M_{j \pm}=
\begin{pmatrix}
\frac{\cos(q_{j}\Delta x\mp\Omega_{j})}{\cos\Omega_{j}} & i\frac{k_{j}}{q_{j}%
}\sin(q_{j}\Delta x) \\
i\frac{k_{j}^{\prime }\sin(q_{j}\Delta x)}{k_{j}\cos\Omega_{j}} & \frac{%
\cos(q_{j}\Delta x\pm\Omega_{j})}{\cos\Omega_{j}})%
\end{pmatrix}%
.
\end{equation}
Here $\Omega_{j}$=$%
arcsin(k_{y}/k_{j}^{\prime })$, $k_{j}$=$(E-V_{j})/\hbar\upsilon _{F}$, $k_{j}^{\prime 2}$=%
$k_{y}^{2}+q_{j}^{2}$, $t^{\prime }$ $\rightarrow$$t^{\prime }/\hbar\upsilon_{F}$, and $q_{j}
$=$sign(k_{j})\sqrt{k_{j}^{2}-k_{y}^2-t^{\prime }k_{j}}$. The entire transfer matrix $X$=$\prod_{j=1}^{2N}M_{j}(w_{j},E,k_{y})$ is for the whole structure. We also get
the reflection coefficient $r$=$r(E,k_{y})$ and the transmission coefficient $t$=$t(E,k_{y})$
\begin{eqnarray}
r=\frac{(k^{\prime
}/k)(x_{22}e^{i\Omega_{0}}-x_{11}e^{i\Omega_{e}})-(k^{\prime
2}x_{12}e^{i(\Omega_{0}+\Omega_{e})}+x_{21}} {(k^{\prime
}/k)(x_{22}e^{-i\Omega_{0}}+x_{11}e^{i\Omega_{e}})-(k^{\prime
2}x_{12}e^{i(\Omega_{e}-\Omega_{0})}-x_{21}}
\end{eqnarray}
\begin{eqnarray}
t=\frac{2(k^{\prime }/k)\cos\Omega_{0}} {(k^{\prime
}/k)(x_{22}e^{-i\Omega_{0}}+x_{11}e^{i\Omega_{e}})-(k^{\prime
2}x_{12}e^{i(\Omega_{e}-\Omega_{0})}-x_{21}}.
\end{eqnarray}
$x_{i,j}(i,j=1,2)$ is the element of total transfer matrix $X$. Meanwhile,
if we define a matrix $Q$ as
\begin{eqnarray}
Q(\Delta x_{j},E,k_{y})=M_{j}(\Delta
x_{j},E,k_{y})\prod_{n=1}^{j-1}M_{n}(w_{n},E,k_{y}),
\end{eqnarray}
here $w_{n}$ is the width in the $n$th potential and the matrix $Q$ denotes
the transport of particles in the $x$ direction. So the wave functions $\varphi_{m}$  at
any coordinate $x_{j-1}+\Delta x_{j}$ %can be given as
\begin{eqnarray}
\lefteqn{\varphi_{m}=(1+r)Q_{m1}+(k'/k)Q_{m2}(e^{i\Omega_{0}}-re^{-i%
\Omega_{0}})}  \notag \\
&&{}+(1+r)Q_{m3}+(k^{\prime }/k)Q_{m4}(e^{i\Omega_{0}}-re^{-i\Omega_{0}})
\end{eqnarray}
where $m$=1,2,3,4 and $Q_{mn}$ are the elements of matrix $Q$.
\begin{figure}[tbp]
\includegraphics[scale=0.375]{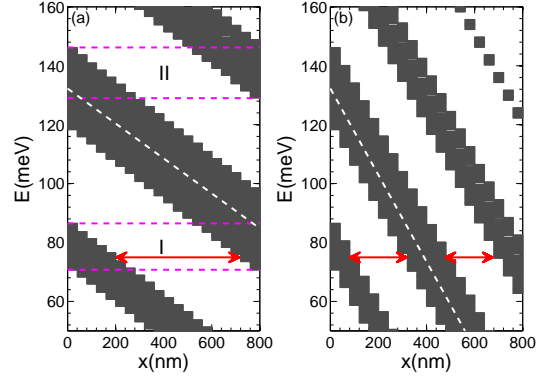} \centering
\caption{(Color online) Trace maps as a function of the depth %in the sample
with (a) $\Delta{V}$=2.5meV, (b) $\Delta{V}$=6meV for %. The other parameters are
N=20, $t^{\prime }$=40meV, $w_{A}$=25nm, $w_{B}$=15nm, $V_{A1}$=150meV, $V_{B1}$=50meV and $k_{y}$=0.}
\label{Fig:fig2}
\end{figure}

%Transfer matrix method can be used
It is conveniently to calculate the band
structures, transmission spectrum and reflection coefficient by transfer matrix method\cite{LGWang2010}.
Fig. \ref{Fig:fig2} represents the trace maps for two situations with the
change of the depth in the system shown as Fig.\ref{Fig:fig1}(a). One can
%clearly
see that the minibands (white) and minigaps (gray) incline under the
external inclined potential, which implies that the band edges depend
linearly on the depth of sample. The trace maps becomes titling, thus the
occurrence of BO becomes possible. In Fig. \ref{Fig:fig2}(a),
the white line in bandgap denotes the central position of zero-$\bar{k}$ gap\cite{CL2013}, and we highlight
two regions called region \uppercase\expandafter{\romannumeral1} and region %
\uppercase\expandafter{\romannumeral2} using two adjacent horizonal lines. EBO usually takes place in these regions, so we call them Bloch zone.
%Taking region \uppercase\expandafter{\romannumeral2} as an example,
%upon being accelerated by the field to the right miniband gaps, particles
%are Bragg reflected and travels to the left. Then it is decelerated by the field until
%reflected by the left band gaps. Then the particle does recipracating motion and the
%so-called Bloch oscillation occur.
Certainly, there shall be a general condition for such
occurrence of BO, which will be discussed latter. Compared with Fig. \ref%
{Fig:fig2}(a), a larger $\Delta{V}$ makes the structure more tilting and the
two Bloch zones overlap. Assuming that an electron with energy of 75meV, as
the red arrows show, it will oscillate in the left band. Because of
the overlap of the Bloch zone, electrons may resonant tunnel from the left
miniband to the right and oscillate in the right band, which is called Zener
tunneling. Thus, Bloch-Zener oscillation takes place. We could have a better
understanding of this phenomenon with transmission spectrum in Fig. \ref%
{Fig:fig3}.
\begin{figure}[tbp]
\includegraphics[scale=0.375]{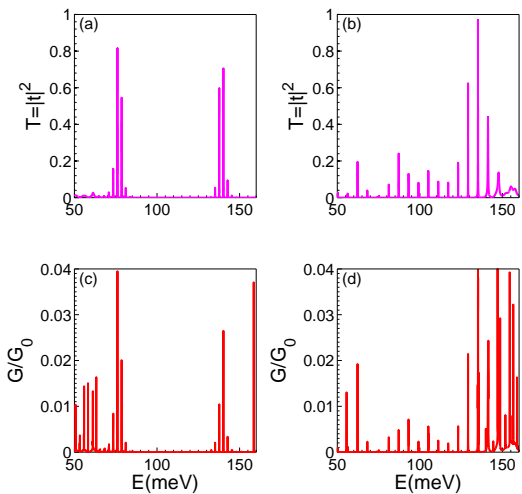} \centering
\caption{(Color online) Transmission spectrum (a) and (b), and conductance $G$ (c) and (d) as a function of Fermi energy. The parameters of (a) and (c), (b) and (d)
are corresponding to Fig. \protect\ref{Fig:fig2}(a) and Fig. \ref{Fig:fig2}(b), respectively.}
\label{Fig:fig3}
\end{figure}

\begin{figure}[tbp]
\includegraphics[scale=0.375]{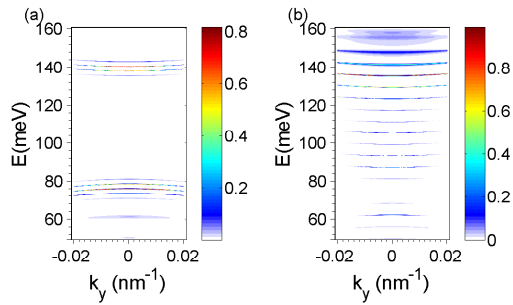} \centering
\caption{(Color online) The distribution of transmission spectrum [(a) and
(b)] with energy and $k_{y}$ corresponding to Figs. \ref{Fig:fig2}(a) and \ref{Fig:fig2}(b).}
\label{Fig:fig4}
\end{figure}
%In Figs. \ref{Fig:fig3}(a) and \ref{Fig:fig3}(b), transmission spectrums $%
%T=|t|^2$ are calculated by Eq. (6) corresponding to Figs. \ref{Fig:fig2}(a)
%and \ref{Fig:fig2}(b), respectively.

Figs. \ref{Fig:fig3}(a) and (b) show the behavior of transmission spectrums.
From Figs. \ref{Fig:fig3}(a), one can
see that there are several transmission peaks equally spaced by $\Delta E_{B}
$, which demonstrates the formation of WSLs, but electrons are totally
reflected when the incident energy is outside of the Bloch zone. Conductance
$G$ is %also
a remarkable transport factor\cite{SDatta1995,TXMa2012,CL2013}, which are plotted in Figs. \ref{Fig:fig3}(c) and (d),
% calculated by $G$=%
%$G_{0}\int_{0}^{\pi/2}T\cos\theta_{0}d\theta_{0}$, $G_{0}$=$2e^{2}m\upsilon
%_{F}L_{y}/\hbar ^{2}$, with $L_{y}$ denoting the width of the graphene
%stripe in the $y$ direction.
corresponding to Figs. \ref{Fig:fig2}(a) and (b), respectively. %One can find conductance
Obviously,  $G$ shows equally
spaced peaks and the positions of these peaks are the same with transmission
spectrum in Figs. \ref{Fig:fig3}(a) and (b). In Figs. \ref{Fig:fig3}(b) and
(d), one can clearly see that Bloch-Zener oscillation occur even though the
BO takes the main part because electrons can transmit at the whole energy
range. The main parameter ruling the dynamics of BO is the oscillation
period $T_{B}=h/\Delta E_{B}$, where $h$ is the Plank constant. %
The $T_{B}$ shown in Figs. \ref{Fig:fig3}(a) and (b) is 1.64ps and 0.69ps,
respectively. Their oscillation frequencies, locate at terahertz.

\begin{figure}[tbp]
\includegraphics[scale=0.375]{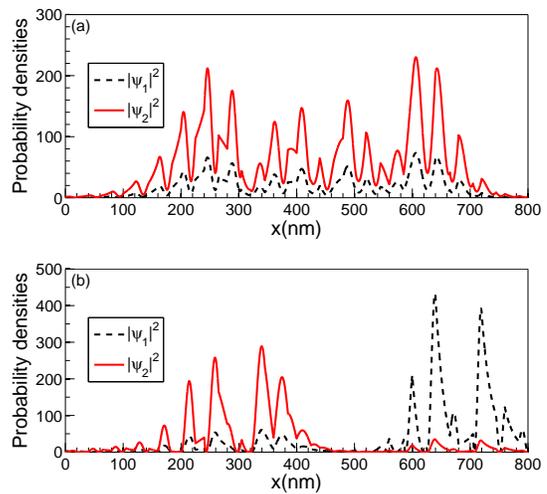} \centering
\caption{(Color online) The evolutions of [(a) and (b)] %probability densities
$|\protect\varphi_{1}|^{2}$ and $|\protect\varphi_{2}|^{2}$ with %the change
of depth corresponding to Figs. \ref{Fig:fig2}(a) and \ref{Fig:fig2}(b) for $E$=76.0639meV (a) $E$=62.2676meV (b) at $k_y=0$.}
\label{Fig:fig5}
\end{figure}

WSLs, with an energy separation $\Delta E_{B}$ between
adjacent resonance, are frequency domain counterpart of BO.
%{\color{blue} I
%am not sure about this sentence: In the fully quantum-mechanical picture},
BO is quantum-interference phenomenon involving Wannier-Stark states\cite%
{HK1996}. As one can see, WSLs are seen as colorized horizonal lines equally
spaced with $\Delta E_{B}$ in Figs. \ref{Fig:fig4}(a) and (b),
and their positions are corresponding to the transmission peaks in Fig. \ref%
{Fig:fig3} greatly. Here WSLs are also direct evidence for the occurrence of
BO. Moreover, the transmission values vary with $k_{y}$ in Fig. \ref%
{Fig:fig4}, which means transmission spectrum is sensitive to the incident
angles. The transmission value takes maximum value when electrons are at normal
incident.

%We turn to study the wave functions defined in Eq. 7. %inside the GSLs calculated by Eq. (7).
%We find that $\varphi_{1}$ and $\varphi_{3}$($\varphi_{2}$ and $\varphi_{4}$) have the same values because of the symmetry
%of Hamiltonian.

In Figs. \ref{Fig:fig5}(a) and (b), we plot
the wave functions defined by Eq. 7, corresponding to Figs. \ref{Fig:fig2}(a) and (b), respectively.
We find that $\varphi_{1}$ and $\varphi_{3}$($\varphi_{2}$ and $\varphi_{4}$) have the same values because of the symmetry of Hamiltonian.
The incident energy of electrons in Fig. \ref{Fig:fig5}(a)% is 76.0639meV, which
matches one of the transmission peaks in
Fig. \ref{Fig:fig3}(a) rather well. %greatly
%(In Fig. \ref{Fig:fig5}(b), it is 62.2676meV).
Fig. \ref{Fig:fig5} (a) demonstrates that wave functions are highly localized %at some special
%regions. In Fig. \ref{Fig:fig5}(a), it is seen
from 100nm to 700nm, which
corresponds to Fig. \ref{Fig:fig2}(a) perfectly. The localized wave functions
means electrons oscillate at this region. %In other words, wave function's
%localization demonstrates the occurrence of BO.
Intriguingly, the wave
functions localize at two parts of the system, see Fig. \ref{Fig:fig5}(b). This can be explained by the effect of Bloch-Zener oscillation. % as Fig. \ref{Fig:fig2}(b).

\begin{figure}[tbp]
\includegraphics[scale=0.375]{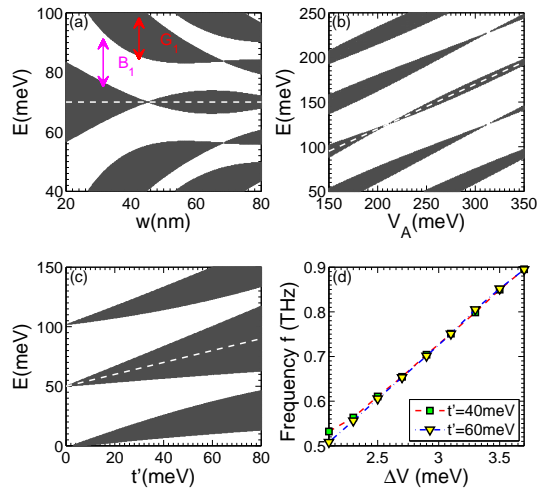} \centering
\caption{(Color online) Dependence of band-gap structure on (a) lattice
constant $w$, (b)barrier potential and (c)$t^{\prime }$. (d) The influence
of potential gradient $\Delta{V}$ and $t^{\prime }$  on $T_{B}$.
}
\label{Fig:fig6}
\end{figure}

\begin{figure}[tbp]
\includegraphics[scale=0.375]{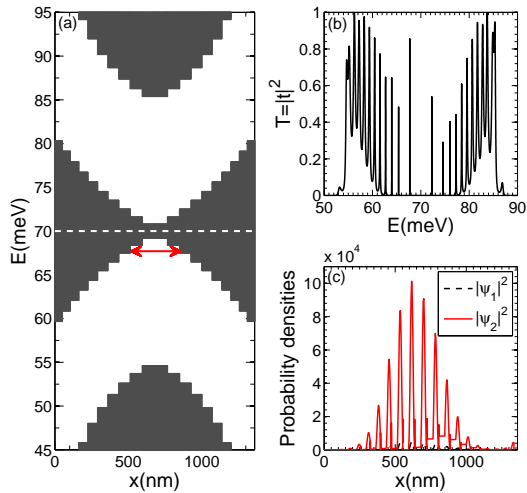} \centering
\caption{(Color online) (a) Band structure correspond to Fig. \protect\ref%
{Fig:fig2} (b), $\Delta{w}=2$nm. (b)Transmission spectrum, and (c)
Distribution of probability density with $E$=67.71meV.}
\label{Fig:fig7}
\end{figure}

Finding a general condition to control the BO is more useful. In Fig. \ref%
{Fig:fig6}(a), we define the zero-$\bar{k}$ gap as the center and label the
gap (G denotes its width) and minibands(B denotes its width) from center in
order. Combining with Fig. \ref{Fig:fig2}, the general condition for
occurrence of BO can be expressed as $B_{1}\leq \Delta{V}\times(N-1)\leq
B_{1}+G_{1}$. For Bloch-Zener oscillation, it is $\Delta{V}\times N\geq
B_{1}+G_{1}$. Thus one can control the occurrence of BO by adjusting the
number of period $N$ or the gradient of potential $\Delta{V}$ . But the
width of minibands and gaps is dominated by several parameters such as
potential values and barrier's width and interlayer coupling.
%As one can see
%in
Fig. \ref{Fig:fig6}(a) shows the width of gaps and minibands highly depend on
the barriers' widths, and
%It decreases and increases periodically with the
%concurrent addition of barriers and wells.
Figs. \ref{Fig:fig6}(b) and \ref{Fig:fig6}(c) show the effect of potential values and $t^{\prime }$, respectively.

The domain parameter in BO mechanic is the frequency %period$T_{B}$,
and we study the
effect of $t^{\prime }$ and $\Delta{V}$ on $T_{B}$ in Fig. \ref{Fig:fig6}%
(d). It is obvious that $T_{B}$ decreases linearly as $\Delta{V}$ increases,
%becomes much larger, %bigger,
which provides a proper way to control the Bloch frequency. A much larger % bigger
$t^{\prime }$ is necessary for the developing of BO as the gaps become larger with a larger $t^{\prime }$, which is helpful for the reflection of electron at the boundary of minibands.

In another system shown as Fig. \ref{Fig:fig1}(b), BO can also emerge. In
the vicinity of zero-$\bar{k}$ gap shown as white horizonal line in Fig. \ref%
{Fig:fig7}(a), system has robust transport properties. The band structure is
symmetric and changes non-linearly with depth of sample, but the zero-$\bar{k%
}$ gap stays at the same position. There is also a Bloch zone at 60-80meV.
Different from the transmission spectrum in Fig. \ref{Fig:fig3}, the
distance between the peaks in Fig. \ref{Fig:fig7}(b) is not the same. Wave
functions localize at some special regions shown as Fig. \ref{Fig:fig7}(c),
which coincides perfectly with the band structure, as the red arrows shown
in Fig. \ref{Fig:fig7}(a).

In summary, EBO and Bloch-Zener oscillation in two
kinds of BLG GSLs are investigated. It is found that terahertz BO and WSLs can be generated in BLG GSLs. In our
systems, the band structures are spatially titling and separated by gaps.
WSLs appear as a series of equally spaced transmission peaks. By changing
the potential's gradient (height or width), BO's period can be
tunable linearly.  Our proposal is in principle possible realized by time-resolved electro-optic technique\cite{TD1994,MH1999,YS2004}.
These results may be important to the applications in
graphene-based electronics\cite{MB2009,GJ2011,LK2012,ED2014}.

%\section{\uppercase\expandafter{\romannumeral4}. conclusion}
%\section{\uppercase\expandafter{\romannumeral4}. conclusion}

%\section{acknowledgments}
T. Ma thanks CAEP for partial financial support. This work is supported by
NSFCs (Grant. Nos. 11104014, 11374034, 11274275, and 61078021),
%Research Fund for the Doctoral Program of Higher Education of China
%20110003120007, SRF for ROCS (SEM),
and the National Basic Research Program of China (Grant Nos. 2011CBA00108,
and 2012CB921602). Hemeng Cheng and Changan Li contributed equally to
this work.

\end{document}